\documentclass[conference]{IEEEtran}
\IEEEoverridecommandlockouts
\usepackage{cite}
\usepackage{amsmath,amssymb,amsfonts}
\usepackage{algorithmic}
\usepackage{graphicx}
\usepackage{textcomp}
\usepackage{xcolor}
\usepackage{url}
\usepackage{enumitem} 

\usepackage{gensymb}
\usepackage{pgfplots}
\usepackage{tikz}
\usepackage{color}
\usepackage{url}

\usepackage{xcolor}
\usepackage{mdwmath}
\usepackage{mdwtab}

\usepackage{lipsum}
\usepackage{capt-of}

\usepackage{caption}
\usepackage{hyperref}
\DeclareFontEncoding{TS1}{}{}
\DeclareFontSubstitution{TS1}{cmr}{m}{n}
\DeclareTextSymbol{\textordmasculine}{TS1}{186}
\DeclareTextSymbolDefault{\textordmasculine}{TS1}

\definecolor{dkgreen}{rgb}{0,0.6,0}
\definecolor{gray}{rgb}{0.5,0.5,0.5}
\definecolor{mauve}{rgb}{0.58,0,0.82}

\colorlet{punct}{red!60!black}
\definecolor{background}{HTML}{EEEEEE}
\definecolor{delim}{RGB}{20,105,176}
\colorlet{numb}{magenta!60!black}

\usepackage{listings,multicol}
\lstdefinelanguage{json}{
	basicstyle=\normalfont\ttfamily,
	numbers=left,
	numberstyle=\scriptsize,
	stepnumber=1,
	numbersep=1pt,
	showstringspaces=false,
	breaklines=true,
	frame=lines,
	framerule=0.5pt,
	xleftmargin=2pt,
	xrightmargin=2pt,
	breakindent=10pt,
	resetmargins=true,
	backgroundcolor=\color{background},
	literate=
	*{0}{{{\color{numb}0}}}{1}
	{1}{{{\color{numb}1}}}{1}
	{2}{{{\color{numb}2}}}{1}
	{3}{{{\color{numb}3}}}{1}
	{4}{{{\color{numb}4}}}{1}
	{5}{{{\color{numb}5}}}{1}
	{6}{{{\color{numb}6}}}{1}
	{7}{{{\color{numb}7}}}{1}
	{8}{{{\color{numb}8}}}{1}
	{9}{{{\color{numb}9}}}{1}
	{:}{{{\color{punct}{:}}}}{1}
	{,}{{{\color{punct}{,}}}}{1}
	{\{}{{{\color{delim}{\{}}}}{1}
	{\}}{{{\color{delim}{\}}}}}{1}
	{[}{{{\color{delim}{[}}}}{1}
	{]}{{{\color{delim}{]}}}}{1},
}
\lstdefinestyle{json}
{
	frame=lrb,         
	belowcaptionskip=-1pt,
	xleftmargin=8pt,
	framexleftmargin=8pt,
	framexrightmargin=5pt,
	framextopmargin=5pt,
	framexbottommargin=5pt,
	framesep=0pt,
	rulesep=0pt,
}
\def\BibTeX{{\rm B\kern-.05em{\sc i\kern-.025em b}\kern-.08em
		T\kern-.1667em\lower.7ex\hbox{E}\kern-.125emX}}
\begin{document}
	
\title{DeepSun: Machine-Learning-as-a-Service for Solar Flare Prediction}
\author{\IEEEauthorblockN{Yasser Abduallah, Jason T. L. Wang, Yang Nie}
		\IEEEauthorblockA{\textit{Department of Computer Science} \\
			\textit{New Jersey Institute of Technology}\\
			University Heights, Newark, NJ 07102, USA} \\
		\and
		\IEEEauthorblockN{Chang Liu, Haimin Wang}
		\IEEEauthorblockA{\textit{Institute for Space Weather Sciences} \\
			\textit{New Jersey Institute of Technology}\\
			University Heights, Newark, NJ 07102, USA} \\		
	}
	
	\maketitle

	\begin{abstract}
		Solar flare prediction plays an important role in understanding and forecasting space weather. 
		The main goal of the Helioseismic and Magnetic Imager (HMI), 
		one of the instruments on NASA's Solar Dynamics Observatory, is to study 
		the origin of solar variability and characterize the Sun's magnetic activity. 
		HMI provides continuous full-disk observations of 
		the solar vector magnetic field with high cadence data that lead to reliable predictive capability; 
		yet, solar flare prediction effort utilizing these data is still limited. 
		In this paper, we present a machine-learning-as-a-service (MLaaS) framework, called DeepSun, 
                    for predicting solar flares on the Web based on HMI's data products. 
		Specifically, we construct training data by utilizing the physical parameters 
		provided by the Space-weather HMI Active Region Patches (SHARP)
		and categorize solar flares into four classes, namely B, C, M, X, 
                     according to the X-ray flare catalogs available at the National Centers for Environmental Information (NCEI).
		Thus, the solar flare prediction problem at hand is essentially a multi-class (i.e., four-class) classification problem.
		The DeepSun system employs several machine learning algorithms to tackle this multi-class prediction problem and 
		provides an application programming interface (API) for remote programming users.
		To our knowledge, DeepSun is the first MLaaS tool capable of predicting solar flares through the Internet.
	\end{abstract}
	
	
	\section{Introduction}\label{sec:intro}
	Solar flares and the often-associated coronal mass ejections (CMEs) 
	highly impact the near-Earth space environment \cite{Liu..Wang..Solar..2017ApJ...843..104L,Liu_2020}. 
	They have the potential to cause catastrophic damage to technology infrastructure \cite{Daglis:2004:EffectSpace}. 
	According to the U.S. National Space Weather Strategy, 
	released by the Space Weather Prediction Center, 
	it is a challenging task to correctly predict
	solar flares and CMEs.
	Recent efforts led by the United States and its partners resulted in substantial progress 
	toward monitoring, prediction, and mitigation plans, but much more effort is still needed. 

	Researches have indicated that the magnetic free energy stored in the corona, 
	quickly discharged by magnetic reconnection,  
	powers solar flares and CMEs \cite{Priest:AA:Magnetic:Nature:SF:2002A&ARv..10..313P} . 
	The process of building the coronal free energy is controlled by the structural evolution of the magnetic field 
	on the photosphere where plasma dominates the process. 
	Observing and measuring the structure and evolution of the photospheric magnetic field 
	can provide valuable information and clues to the triggering mechanisms of 
	flares and CMEs.
	There are many physical properties or parameters, 
	as we will discuss later in the paper,
	that characterize the static photospheric magnetic field, such as 
	integrated Lorentz force, magnetic helicity injection, unsigned magnetic flux, 
	vertical electric currents, magnetic shear and gradient, 
	and magnetic energy dissipation.
	
	Researchers spent significant efforts attempting to understand 
	the physical relationship between flare productivity and non-potentiality of 
	active regions (ARs) as specified by the physical parameters. 
	This led researchers to use different methods to predict flares 
	that are not based on physical models, 
	but rather based on statistical modeling and machine learning  
	\cite{Barnes:2016ApJ...829...89B}. 
	Machine learning gives computer programs
	the ability to learn from data and progressively improve performance. 
	It uses input data, also called training data, and learns hidden insights in the training data 
	to build a predictive model that will be used later to make predictions on unseen test data.

	In our previous work \cite{Liu..Wang..Solar..2017ApJ...843..104L}, we
	reported the results of solar flare prediction using the random forests (RF) algorithm \cite{RandomForestCART1984}. 
           We constructed a database of solar flare events 
         using the physical parameters provided by the Space-weather HMI Active Region Patches (SHARP),
                and categorized solar flares into four different classes, namely B, C, M, X,
              based on the X-ray flare catalogs available at the National Centers for Environmental Information (NCEI).
            Flares in the B class have the smallest magnitude while flares in the X class have the largest magnitude.
              We used the RF algorithm and the physical parameters or features
	to perform multi-class classification of solar flares,
         predicting the occurrence of a certain class of flares in a given active region (AR) within 24 hours.
          Our experimental results demonstrated the good performance of the RF algorithm.
	
	In this paper, we extend our previous work in \cite{Liu..Wang..Solar..2017ApJ...843..104L} by considering 
	two additional multi-class classification algorithms: 
	multilayer perceptrons (MLP) 
	and extreme learning machines (ELM).
	We implement these algorithms into a machine-learning-as-a-service (MLaaS) framework,
	called DeepSun, which allows scientists to perform multi-class flare prediction on the Internet.
	Specifically, our work here makes two contributions. 
	\begin{enumerate}
		\item 
		We develop an ensemble method for multi-class flare prediction that performs better than
		the existing machine learning algorithms including RF, MLP and ELM
		according to our experimental study.	
		\item 
		We design and implement DeepSun, which is the first MLaaS system of its kind
		for solar flare prediction. 
	\end{enumerate}
	The rest of this paper is organized as follows. 
	Section \ref{sec:13features} describes the data and the SHARP predictive parameters used in this study.
           Section \ref{sec:ml} describes the machine learning algorithms
	employed by DeepSun. 
	Section \ref{sec:results} evaluates the performance of these machine learning algorithms.
	Section \ref{sec:methodology} details
	the design and implementation of the DeepSun framework.
           Section \ref{sec:relatedwork} surveys related work and compares DeepSun with existing services computing systems.
	Section \ref{sec:discussionfuturework} concludes the paper and 
	points out some directions for future research.
	
	\section{Data and Physical Parameters} 
          \label{sec:13features}
          In 2012, SHARP data were released.
          The main goal of the SHARP data was to facilitate AR (active region) event forecasting \cite{Bobra:2014SoPh..289.3549B}. 
	These data are available in the Joint Science Operations Center (JSOC) (\url{http://jsoc.stanford.edu/})
	as hmi.sharp series which include magnetic measures and parameters
          for many ARs. 
	In 2014, another data series, cgem.Lorentz, were produced based on the SHARP data. 
	This series include the Lorentz force estimations.
	 The main goal of this series was to help diagnose the dynamic process of ARs. 
          Bobra {\it et al.} \cite{Bobra:2014SoPh..289.3549B} considered 25 physical parameters in the 
	SHARP datasets that characterize the AR magnetic field properties. 
	The authors used a univariate feature selection method to score the 25 parameters, and
	suggested that the top 13 out of the 25 parameters be used as predictors for flare activity. 
	Table \ref{tab:13features} summarizes these 13 important parameters and their descriptions.
	More details about the 13 physical parameters can be found in \cite{Liu..Wang..Solar..2017ApJ...843..104L}. 
	
	\begin{table}[t]
		\renewcommand{\arraystretch}{1.3}
		\caption{13 Important SHARP Physical Parameters} 
		\label{tab:13features}
		\centering
		\begin{tabular}{l|l}
			\hline	
			\bfseries Parameter & \bfseries Description\\
			\hline
			ABSNJZH & Absolute value of the net current helicity \\
			AREA\_ACR & Area of strong field pixels in the active region \\
			EPSZ & Sum of z-component of normalized Lorentz force \\
			MEANPOT & Mean photospheric magnetic free energy \\
			R\_VALUE & Sum of flux near polarity inversion line \\
			SAVNCPP & Sum of the modulus of the net current per polarity \\
			SHRGT45 & Fraction of area with shear $>$ 45$\degree$ \\
			TOTBSQ & Total magnitude of Lorentz force \\
			TOTFZ & Sum of z-component of Lorentz force \\
			TOTPOT & Total photospheric magnetic free energy density \\
			TOTUSJH & Total unsigned current helicity \\
			TOTUSJZ & Total unsigned vertical current \\
			USFLUX & Total unsigned flux \\
			\hline
		\end{tabular}
	\end{table}

\begin{figure*}
		\centering
		\includegraphics[width=2.0\columnwidth]{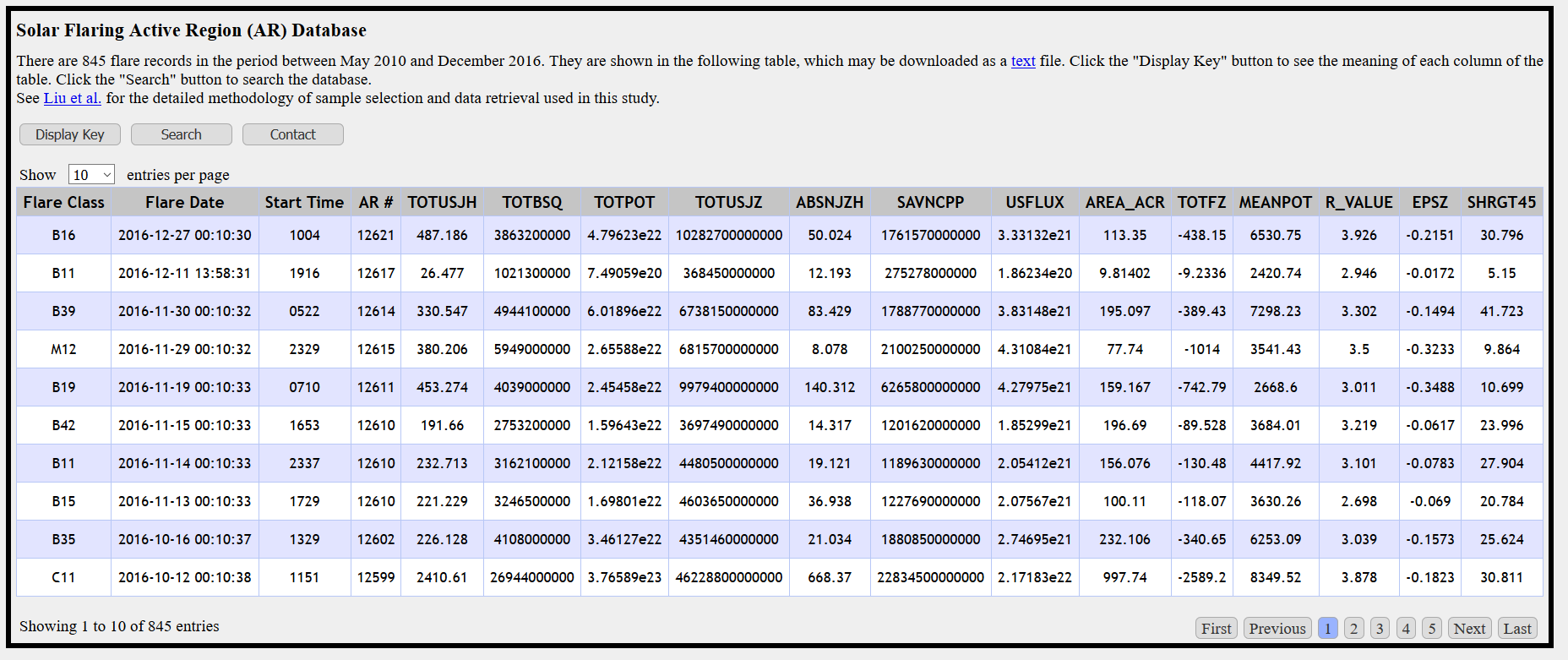}
		\caption{Screenshot showing our online flare database.}
		\label{fig:flareDB}
\end{figure*}

	 We constructed a database based on the SHARP parameters extracted from the solar images that are available at the Joint Science Operations
	Center (JSOC) and the X-ray flare catalogs provided by the National Centers for Environmental Information (NCEI) \cite{Liu..Wang..Solar..2017ApJ...843..104L}.
	We considered the period between May 2010 and December 2016.
	There are 845 flares in this period, among which
	128 flares are of class B, 552 flares are of class C, 142 flares are of class M, and 23 flares are of class X. 
	These 845 flares come from 472 active regions (ARs).
	 The duration of a flare ranges from several minutes to hours.
	The duration of an AR ranges from several minutes to days.
	Table \ref{tab:classwithregisions} summarizes the flare information.

          We created and stored 845 corresponding data samples in our database, shown in Figure \ref{fig:flareDB} and accessible at
       \url{https://nature.njit.edu/spacesoft/Flare-Predict/},
          where each data sample 
	contains values of the 13 physical parameters or features listed in Table \ref{tab:13features}.
        The two digits following a class label (B, C, M, X) are ignored in performing flare prediction.
        The time point of a data sample is the beginning time (00:00:01 early morning) 
        of the start date of a flare and the label of the data sample is the class which the flare belongs to.
       These labeled data samples are used to train the DeepSun system.

	\begin{table}[!t]
		\renewcommand{\arraystretch}{1.3}
		\caption{Numbers of Flares and Active Regions per Solar Flare Class}
		\label{tab:classwithregisions}
		\centering
		\begin{tabular}{c|c|c}
			\hline
			\bfseries Flare Class & \bfseries Number of Flares & \bfseries Number of ARs\\
			\hline
			B & 128 & 88 \\
			C & 552 & 281 \\
			M & 142 & 88\\
			X & 23 & 15\\
			\hline
		\end{tabular}
	\end{table}
	
	\section{Machine Learning Algorithms}\label{sec:ml}
	DeepSun employs three machine learning algorithms for flare prediction: 
          random forests (RF) \cite{RandomForestCART1984},
         multilayer perceptrons (MLP)  \cite{ThePerceptron..book..1958,ANN:DBLP:conf/ann/1995} and
        extreme learning machines (ELM) \cite{SLNs:Convex:DBLP:journals/ijon/HuangC07, SLNs:Enhan:DBLP:journals/ijon/Huang008}.
       RF is a tree-based algorithm comprised of multiple binary classification and regression trees (CART) while both MLP and ELM are 
     feed-forward artificial neural networks.
       All the three algorithms are well suited for multi-class classification.
      In addition, we develop an ensemble (ENS) algorithm, which works by taking the majority vote of RF, MLP and ELM.
     If there is no majority vote for a test data sample, ENS outputs ``no verdict'' for the test data sample.

\section{Performance Evaluation}\label{sec:results}
We conducted a series of experiments to evaluate the performance of the machine learning algorithms presented in Section \ref{sec:ml}
using the database described in Section \ref{sec:13features}.
To avoid bias and to keep the data as balanced as possible, we created 100 csv (comma separated values) datasets of which 
	each dataset included all B, M, and X classes and randomly selected 142 data samples of class C. 
We used 10-fold cross validation in which
 for each data set, we randomly formed 10-fold partitions using the 
KFold function provided by the scikit-learn library in Python \cite{scikit-learn}.
Each machine learning algorithm was trained by nine of the 10-folds, and the 10th fold was used for testing. 
To overcome errors associated with cross validation, we repeated the procedure 100 times for each of the 100 datasets
 that resulted in 10000 iterations to produce the final result.

	We converted the multiple-class classification problem at hand into four binary classification problems for the four classes B, C, M, X respectively.
	For example, consider the binary classification problem for class B.
	Here, we say a data sample is positive if it is in class B, or negative if it is not in class B, i.e., it is in class C, M or X.
	We define TP (true positive), FP (false positive), TN (true negative), FN (false negative) as follows.
	TP is a data sample where an algorithm predicts the data sample to be positive and the data sample is indeed positive.
	FP is a data sample where the algorithm predicts the data sample to be positive while the data sample is actually negative.
	TN is a data sample where the algorithm predicts the data sample to be negative and the data sample is indeed negative.
	FN is a data sample where the algorithm predicts the data sample to be negative while the data sample is actually positive.
	We also use TP (FP, TN, FN respectively) to represent the number of true positives (false positives, true negatives, false negatives respectively).
 	Because we are tackling an imbalanced classification problem (see Table \ref{tab:classwithregisions}),
         we adopt two performance metrics, balanced accuracy (BACC) and true skill statistics (TSS), where
	BACC is defined as follows: 
	\begin{eqnarray*}
		\mbox{BACC} = \frac{1}{2}\bigg( \frac{\mbox{TP}}{\mbox{TP} + \mbox{FN}} + \frac{\mbox{TN}}{\mbox{TN} + \mbox{FP}}\bigg)
	\end{eqnarray*}
	and TSS is defined as follows: 
	\begin{eqnarray*}
		\mbox{TSS} = \frac{\mbox{TP}}{\mbox{TP} + \mbox{FN}} - \frac{\mbox{FP}}{\mbox{TN} + \mbox{FP}}
	\end{eqnarray*}

	BACC and TSS are well suited for imbalanced classification of solar eruptions \cite{LSTMSolarFlaresLiu_2019,Liu_2020,Liu..Wang..Solar..2017ApJ...843..104L}.
	We obtain BACC and TSS for each binary classification problem.
	There are four binary classification problems. 
	We then calculate the average of the BACC and TSS values
	obtained from the four classification problems, and
	use the average as the result for the multi-class classification problem.

\begin{figure*}[!h]
		\centering
		\includegraphics[width=1.2\columnwidth]{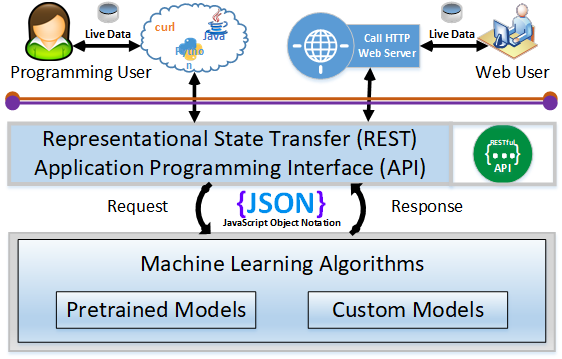}
                     \caption{Overview of DeepSun.}
		\label{fig:contextarch}
\end{figure*}

We implemented the machine learning algorithms in Python leveraging the scikit-learn packages \cite{scikit-learn}. 
Each algorithm has different optimization parameters to 
be tuned based on the training and test datasets. 
We used random forests (RF) composed of 500 to 1000 trees and set the number of features to six to find the best node split. 
For multilayer perceptrons (MLP) and extreme learning machines (ELM), we set the number of hidden layers to 200.
These parameter values were chosen to maximize TSS values.

Table \ref{tab:resultcomparision} compares the BACC and TSS values of the machine learning algorithms at hand
for each binary classification problem and for the overall multi-class classification problem
where the highest performance metric values are highlighted in boldface.
It can be seen from the table that the proposed ENS algorithm is better than the existing algorithms RF, MLP and ELM.
However, all the four algorithms perform poorly in predicting X-class flares.
This happens probably because the X class has much fewer flares than the other classes.
Overall, there were approximately less than 2\% data samples receiving ``no verdict.''

\begin{table}[!h]
		\renewcommand{\arraystretch}{1.3}
		\caption{Flare Prediction Results Using 13 SHARP Parameters and Four Machine Learning Algorithms}
		\label{tab:resultcomparision}
		\centering
		\begin{tabular}{l|ccccc}
			\hline
                                & \bfseries Class B & \bfseries Class C & \bfseries Class M  & \bfseries Class X &\bfseries Average\\
			\hline		
                                BACC & & & & &\\
			ENS  & \bfseries 0.871 & \bfseries0.691 & \bfseries0.790 & \bfseries0.670 &\bfseries 0.756\\
			RF  & 0.834 & 0.663 & 0.749 & 0.645 &0.723\\
			MLP & 0.818&0.659&0.757&0.599 &0.708\\
			ELM & 0.791&0.641&0.721&0.608& 0.690\\
			\hline	
                                TSS & & & &\\
			ENS & \bfseries 0.745 & \bfseries 0.380 & \bfseries 0.551& \bfseries 0.362 & \bfseries 0.507\\		
			RF & 0.708 & 0.378 & 0.537 & 0.330 & 0.488\\			
			MLP & 0.661 & 0.285 & 0.526 & 0.010 & 0.371\\
			ELM &0.618 & 0.296 & 0.446 & 0.227 & 0.397\\
			\hline
		\end{tabular}
\end{table}	

	\section{The DeepSun Framework} \label{sec:methodology}
	\subsection{ System Design}
The four machine learning algorithms (ENS, RF, MLP, ELM) have been implemented into our DeepSun system
where the algorithms are used as a back-end, also known as the server-side, 
engine for the machine-learning-as-a-service (MLaaS) platform.
Fig. \ref{fig:contextarch} presents the overall contextual architecture of the DeepSun framework. 
The system supports two different types of users: Web and programming.
The Web user invokes the service by accessing a graphical user interface (GUI) to perform flare predictions. 
The programming user can use any programming language that supports HTTP requests, 
such as Java, C++, Python, Node.js, JavaScript modules in React or other frameworks to perform flare predictions.

MLaaS is a representational state transfer (REST) application programming interface (API) 
that supports JSON (JavaScript Object Notation) formatted payloads in the request and response. 
JSON is a plain-text and lightweight data-interchange format. 
It is structured with attributes and values in an easy way for humans to read and write. 
JSON is language independent but it is easy to parse; therefore almost every programming language supports it.
The request transmits the user's data from the front-end to the back-end and must include well defined JSON formatted test data to predict or 
training data to create a predictive model. 
The response transmits the result from the back-end to the front-end, which is a well formatted prediction result or the predictive model identifier. 
Here, the front-end means the client-side that can be a Web-designed interface for the Web user or a program for the programming user. 

\subsection{System Implementation}\label{sec:deepsun}
\begin{figure}[h]
		\centering
		\includegraphics[width=1\columnwidth]{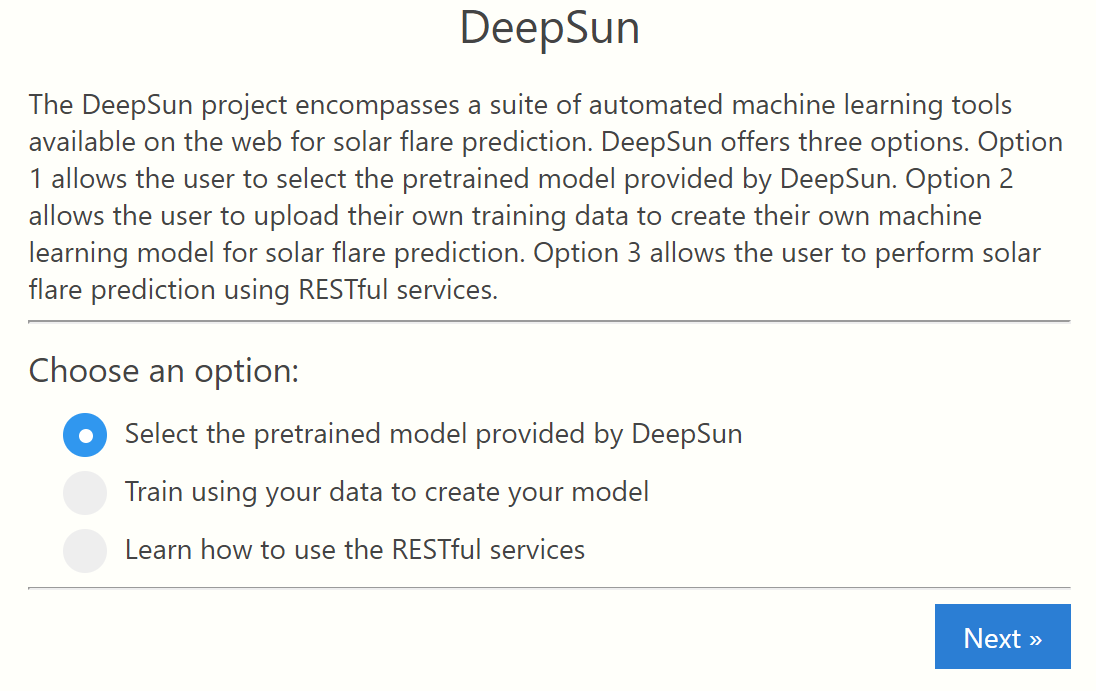}
		\caption{Screenshot showing the home page of DeepSun.}
		\label{fig:deepsunhomepage}
	\end{figure}

When a user visits DeepSun's home page, 
the user sees three options.
Option 1 allows the user to select the pretrained models provided by DeepSun. 
Option 2 allows the user to upload his/her own training data to create his/her own machine learning models for solar flare prediction. 
Option 3 allows the user to perform solar flare prediction using RESTful services.
Fig. \ref{fig:deepsunhomepage} shows DeepSun's home page, 
which can be accessed at
\url{https://nature.njit.edu/spacesoft/DeepSun/}.
	
          \begin{figure*}
		\centering
		\includegraphics[width=1.8\columnwidth]{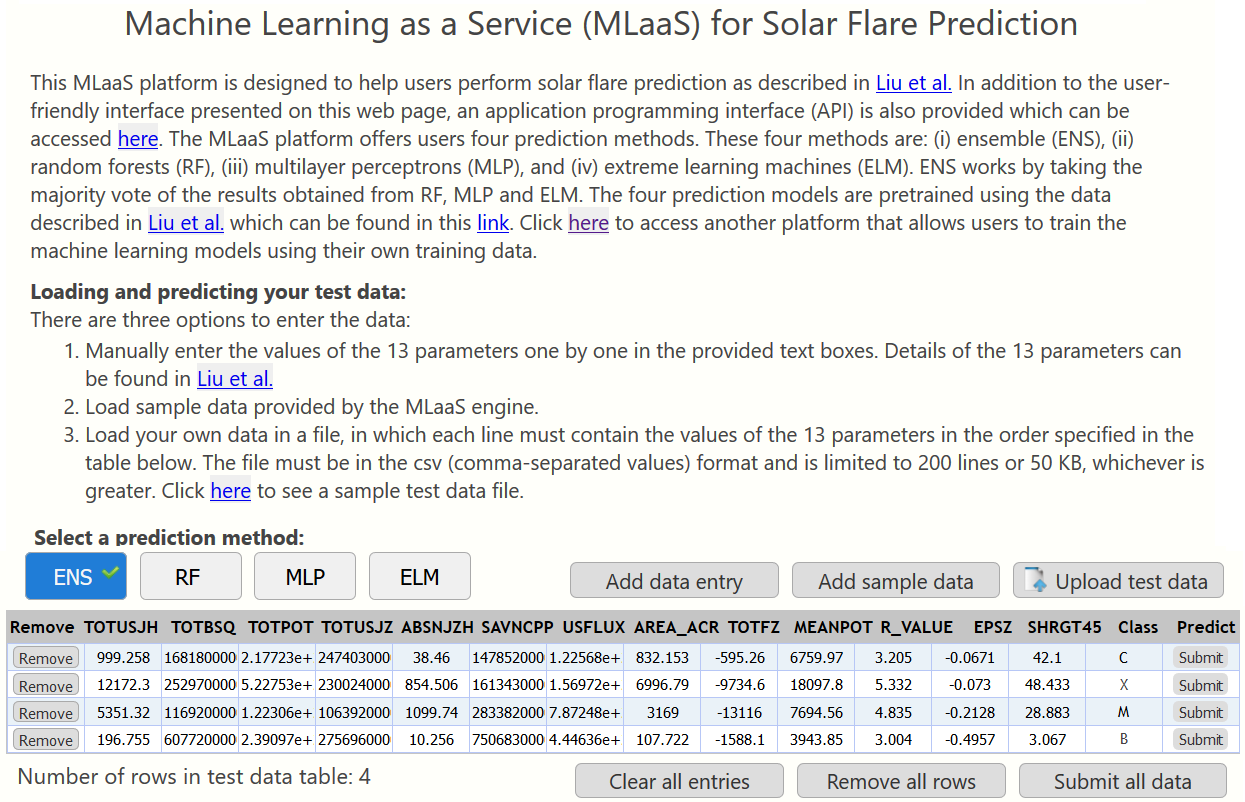}
		\caption{Screenshot showing the webpage with pretrained models of DeepSun.}
		\label{fig:pretrained-model}
	\end{figure*}
	
\subsubsection{Pretrained Models in DeepSun}\label{sec:deesun:pretrained}
The pretrained models are ready-to-use models that were created using the database described in Section \ref{sec:13features}.
With the pretrained models, a user has multiple options to load test data samples containing the 13 physical parameters or features listed in Table \ref{tab:13features}: 
(1) Manually enter the data samples with values of the 13 physical parameters one by one in the provided text boxes.
(2) Load sample data provided by the DeepSun engine. 
(3) Load the user's own data in a file, in which each line contains the values of the 13 physical parameters. 
The user may invoke the services to predict all the loaded, or entered, test data at once or make predictions one by one. 
Fig. \ref{fig:pretrained-model} shows the webpage of pretrained models on which four predictions were made using the ENS algorithm.

\subsubsection{Custom Models in DeepSun} \label{sec:deepsun:custommodel}
\begin{figure*}
		\centering
		\includegraphics[width=1.8\columnwidth]{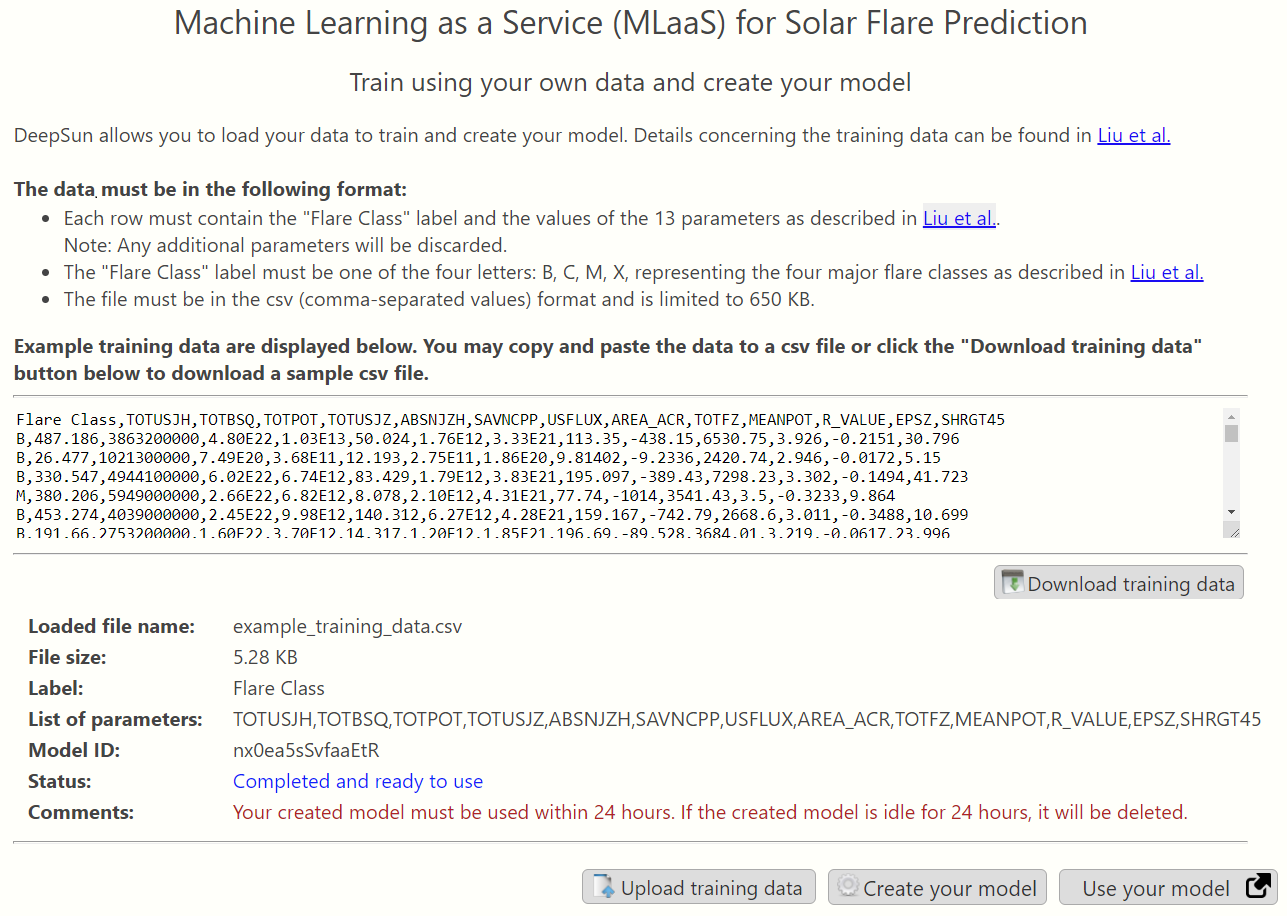}
		\caption{Screenshot showing the webpage with custom models of DeepSun.}
		\label{fig:custom-model}
	\end{figure*}

DeepSun allows the user to load his/her data to train and create his/her custom model to predict solar flares. 
The training data are saved in a file meeting DeepSun's format requirement.
When the user creates a custom model, a model identifier (id) is assigned to the current session. 
If the created model is idle for 24 hours, it will be deleted.
Once the model is ready, the user goes to the DeepSun's graphical user interface with the assigned model id to 
perform flare predictions as done with the pretrained models.
The model id is used to distinguish between the custom models and pretrained models.
Fig. \ref{fig:custom-model} shows the webpage of custom models with example training data displayed.	
	
	\subsubsection{RESTful API for DeepSun} \label{sec:deepsun:restuflapi}
	The RESTful API is designed to help the programming user perform solar flare predictions using the pretrained or custom models. 
	The API supports the POST request to predict solar flare occurrence or create a custom model, 
          and the GET request to get a random data sample from our training database. 
	The interface supports JSON formatted strings for requests' body and their results.
	The interface also supports two different debug levels; they are  
            (i) INFO which is the default debug mode and (ii) DEBUG to return additional data with the result.

	The return result from the POST request is a JSON object including the predicted solar flare occurrence and its class. 
          Each test data sample is associated with a JSON object that includes two attributes. 
        One attribute is ``fcnumber'' which is the numerical representation for the solar flare class
	where we use ``1" (``2", ``3", ``4" respectively) to represent class B (C, M, X respectively).
       The other attribute is "fcname" which is the solar flare class name. 

\begin{figure*}
		\centering
		\includegraphics[width=1.8\columnwidth]{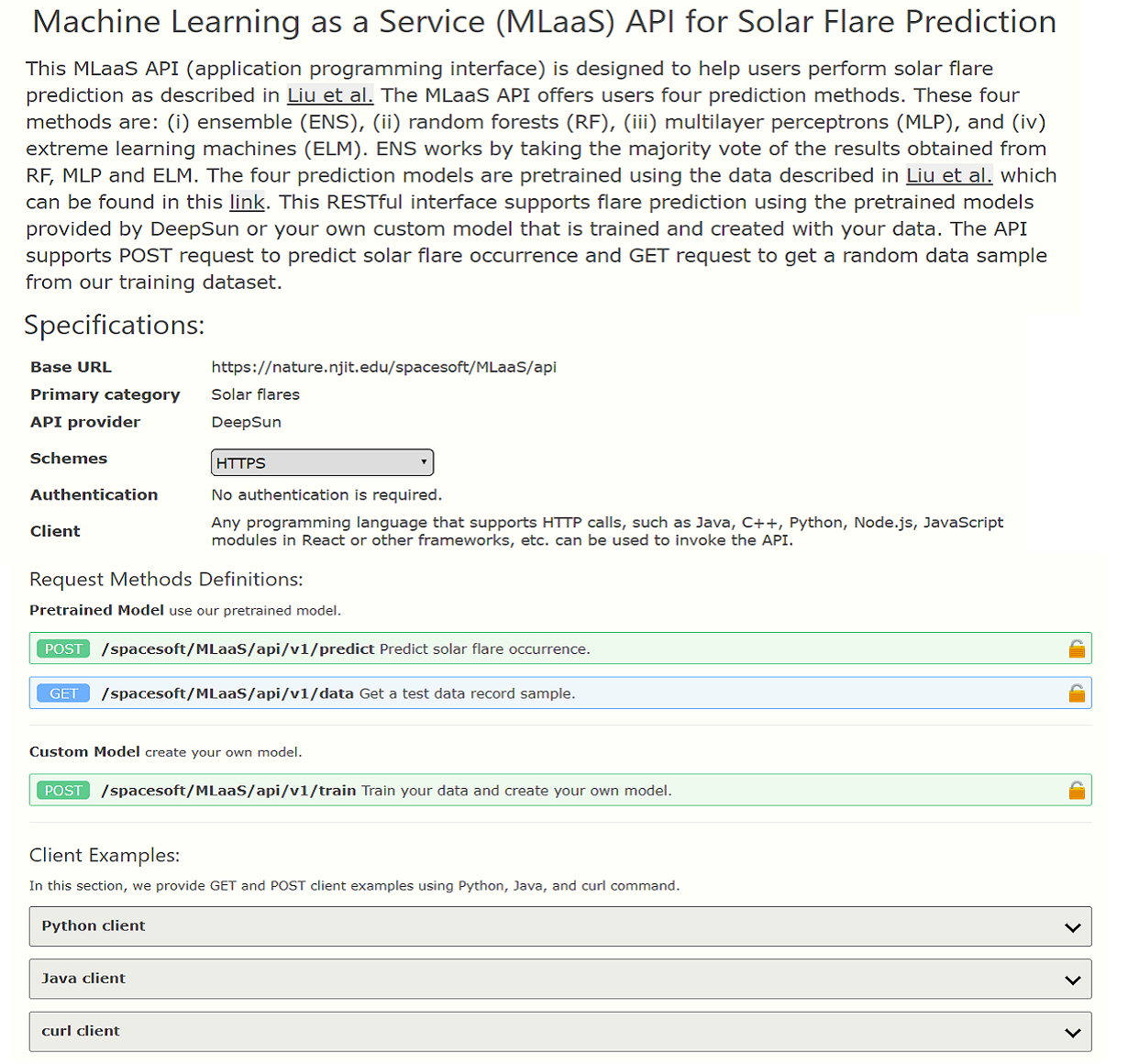}
		\caption{Screenshot showing the RESTful API page of DeepSun.}
		\label{fig:apihomepage}
	\end{figure*}
	
In addition, the RESTful API uses the POST request to create a custom model. 
The body of the request must be JSON formatted strings for an array of JSON objects. 
 Each object must contain the 13 physical parameters and its flare class label where the label must be one of B, C, M, X.
The return result of this POST request is a JSON object that contains the custom model identifier (id) which can be used for flare prediction. 
The custom model includes all the four algorithms (ENS, RF, MLP, ELM).
Since the API is a RESTful interface, any programming language that supports HTTP calls,
 such as Java, C++, Python, Node.js, JavaScript modules in React or other frameworks
 can be used to invoke the API.
Fig. \ref{fig:apihomepage} shows the RESTful API page on which the definitions of the available methods and client examples are displayed. 
	
\section{Related Work} \label{sec:relatedwork}
There are two groups of work that are closely related to ours.
The first group is concerned with solar flare forecasting.
Many studies in this group used parameters derived from the line-of-sight (LOS) component of the photospheric magnetic field 
and produced probability outputs for the occurrence of a certain magnitude flare in a time period \cite{Liu..Wang..Solar..2017ApJ...843..104L}. 
         Some researchers \cite{Gallagher:2002SoPh..209..171G} 
	used sunspot classification and Poisson statistics to 
	provide probabilities for an active region (AR) to produce flares with different magnitudes within 24 hours. 
	Song {\it et al.} \cite{Song:2009SoPh..254..101S} used three LOS magnetic parameters 
	together with the ordinal logistic regression (OLR) method to predict the probabilities of a one-day flare. 
	Bloomfield {\it et al.} \cite{Bloomfield:TSS:Recommend:2012ApJ...747L..41B} suggested that 
	the prediction probabilities should be converted into a binary (i.e., yes-or-no) forecast
	before they can be translated as flare-imminent or flare-quiet. 
	Following this suggestion, Yuan {\it et al.} \cite{Yuan:2010RAA....10..785Y} 
	employed support vector machines (SVMs) to obtain a clear true or false flare prediction for different flare classes.

	On the other hand, the full vector data provide more information 
	about the photospheric magnetic field structure compared to the LOS field. 
	This type of information may provide better flare prediction performance, 
	but due to the limitation imposed by ground-based vector magnetic field observations, 
	the work on flare forecast is limited. 
	For example, Leka and Barnes \cite{Leka:Barnes:2003ApJ...595.1296L} 
	used a small sample of vector magnetograms from the Mees Solar Observatory 
	and applied a discriminant analysis to differentiate between flare-producing and flare-quiet ARs within few hours. 
	The authors later extended their work and 
	used a larger number of samples with a 24-hour prediction window 
	on producing probabilistic forecasts \cite{Barnes:2007SpWea...5.9002B}.

	Since May 2010, the Helioseismic and Magnetic Imager (HMI) onboard the Solar Dynamics Observatory (SDO) 
  	\cite{Bobora:2015ApJ...798..135B} 
	has been producing high quality photospheric vector magnetograms 
	with high-cadence and full-disk coverage data. 
	Using these data,  Bobra and Couvidat \cite{Bobora:2015ApJ...798..135B} 
	calculated a number of magnetic parameters for each AR. 
	They selected 13 from all the available parameters and 
	achieved good prediction performance using an SVM method for flares greater than M1.0 class. 
	Nishizuka {\it et al.} \cite{Nishizuka:2017ApJ...835..156N} applied a number of machine learning algorithms to HMI data 
	and produced prediction models for $\ge$M and X-class flares with reasonably high performance. 
          More recently, we employed a long short-term memory network for flare prediction \cite{LSTMSolarFlaresLiu_2019}.

The second group of related work is concerned with services computing.	
Benmerar {\it et al.} \cite{BrainMRI..Benmerar.e.Al..2017} developed a brain diffusion MRI (magnetic resonance imaging) application
to overcome the SaaS (software-as-a-service) limitations caused by intensive computation. 
The application provides APIs that tackle browser paradigms to reduce the parallel computation rendered in the client side of the browser. 

Wu {\it et al.} \cite{XCheckCrossBrowser..Wu.et.al..2018} 
developed an automated testing technique to detect cross-browser compatibility issues 
so that they can be fixed. 
These cross-browser issues cause problems for an organization to create JavaScript web applications. 
The authors employed an existing record-and-play technique, Mugshot \cite{Mugshot..Mickens.et.al..2010}, 
to design an incremental cross-browser incompatibility algorithm. 
The system starts off by injecting the record library into the browsers, collects traces and events 
to be replayed, and runs the detection algorithm to find different types of incompatibilities among the browsers. 

Song {\it et al.} \cite{ITServiceClassification..Song.et.all..2012} presented a machine learning algorithm for IT support services 
to automate the problem determination and classification, and also find the root cause of a problem. 
The algorithm is an on-line perceptron that learns about the user's problems from the data that were generated from logs and monitoring information across different systems. 
The algorithm then categorizes the problems by finding the actual root cause from what it learned from the data. 
The algorithm employs an incremental learning technique and is able to automatically adjust the classifier parameters. 
	
Li {\it et al.} \cite{APIDocumentationRecommendation..Li.et.al..2018} described a new software documentation recommendation methodology 
that adopts a learn-to-rank (LTR) technique.
LTR is an application of supervised and semi-supervised machine learning techniques. 
Their strategy combines the social context from a questions-and-answers online system 
and the content of official software documentation to build the LTR model to provide accurate and relevant software documentation recommendations. 
Their experimental results showed that this approach outperforms traditional code search engines including the Google search engine.
	
Our DeepSun system differs from the above works in two ways.
First, DeepSun provides services dedicated to solar flare prediction,
which has not been addressed by the existing services computing systems.
Second, in the solar flare forecasting area, DeepSun is the
first MLaaS system, to our knowledge, that allows scientists to perform
multi-class flare prediction through the Internet.
	
\section{Conclusions and Future Work} \label{sec:discussionfuturework}
	
We present a machine-learning-as-a-service framework (DeepSun) for solar flare prediction.
This framework provides two interfaces: a web server where the user enters the information through a graphical interface 
and a programmable interface that can be used by any RESTful client.
DeepSun employs three existing machine learning algorithms, namely
random forests (RF), multilayer perceptrons (MLP), extreme learning machines (ELM), 
and an ensemble algorithm (ENS) that combines the three machine learning algorithms.
Our experimental results demonstrated the good performance of the ensemble algorithm
 and its superiority over the three existing machine learning algorithms.

In the current work we focus on data samples composed of SHARP physical parameters.
We collect 845 data samples belonging to four flare classes: B, C, M, and X across 472 active regions.
In addition, the Helioseismic Magnetic Imager (HMI) aboard the Solar Dynamics Observatory (SDO)
produces continuous full-disk observations (solar images).
In future work we plan to incorporate these HMI images into our DeepSun framework and
extend our previously developed deep learning techniques \cite{LXW20,HuTPW18,HuTW20}
to directly process the images.
We also plan to combine our recently developed deep learning algorithms using 
the SHARP parameters  \cite{LSTMSolarFlaresLiu_2019}
with the image-based techniques and machine learning algorithms described in this paper
for more accurate solar flare prediction.


\end{document}